\documentclass[10pt]{article}
\usepackage{geometry}
\usepackage{cite}
\usepackage{amssymb,amsmath}
\usepackage{wrapfig}
\usepackage{graphicx}
\usepackage[small]{caption}
\usepackage[titletoc,title]{appendix}
\usepackage{authblk}
\usepackage{abstract}

\newtheorem{lemma}{Lemma}
\newtheorem{conjecture}{Conjecture}
\newtheorem{corollary}{Corollary}

\begin{document} 
\title{Conspiracy in ontological models \textendash  \hspace{2pt} $\lambda$-sufficiency and measurement contextuality}
\author{Yiruo Lin \thanks{yiruolin@tamu.edu, linyiruo1@gmail.com}}
\affil{\textit{Department of Computer Science \& Engineering, Texas A\&M University}}
\date{}
\maketitle

\begin{abstract}
\normalsize
The conspiracy of ontic states responding to measurements contextually to comply with noncontextual quantum mechanical probabilities is analyzed for general ontological models. A general physical picture of ontological space structure and how ontic states are disturbed by measurement contexts is presented. A common assumption in ontological models called $\lambda$-sufficiency is analyzed and argued to be inconsistent with measurement contextuality for ontological models with certain symmetry in ontic space with respect to quantum states and measurements. 
\end{abstract}

\section{Introduction}

Understanding the underlying reality behind quantum mechanical formalism has been a long-standing effort that is also of great interest to the current quantum information science. A major motivation behind the research programme is to gain an intuitive physical picture absent in the abstract mathematical formalism of quantum mechanics. Surprisingly and quite remarkably, aided with certain assumptions in which information plays a crucial role classical notion of physics can explain a large part of nonclassical phenomena commonly regarded as typical quantum phenomena \cite{Spekkens_epi},\cite{Spekkens and etc}. However, contextuality and nonlocality as spacetime manifestation of the contextuality remain major puzzles and in this sense can be regarded as true quantum phenomena beyond our current intuitive picture of the physical world. \\

Ontological approach provides a general mathematical framework in which the putative realistic structure underlying quantum mechanics can be analyzed. In the framework, physically realistic states (called ontic states) form an ontological space and a quantum state (pure or mixed) can be regarded as a statistical mixture of the underlying ontic states. The response of an ontic state to projective/sharp observations is typically regarded as deterministic (but not in general predetermined due to contextuality) as it is defined to provide the complete physical property of an isolated system. Of course, there are scenarios where the deterministic feature of an ontic state no longer holds. The underlying physical world can be probabilistic. Another possibility is such that its response to experimental probes depends on the detailed microscopic probe configurations so that at the coarse-grained level, the response is probabilistic due to lack of knowledge of experimental interactions (it is categorized as microscopic deterministic in \cite{Rudolph_Harrigan} in contrast to macroscopic deterministic where the response is indifferent to microscopic details). Assuming determinacy, the ontic state response to a projective measurement depends on the context in which the measurement is performed along with the rest of the measurements, well known as Bell-Kochen-Specker (BKS) theorem\cite{Bell},\cite{KS}. A relevant generalization of BKS theorem is given by Spekkens\cite{Spekkens_general_context} where general preparation and unsharp measurements (POVM) are included and it is shown that noncontextuality can not hold for both preparation and measurement. \\

Measurement contextuality enforces disturbance on ontic states under measurement, verifying the orthodox view that measurement inevitably perturbs the microscopic states. Nevertheless, it is contrived how the measurement contextuality of individual ontic state can always comply with the statistical prediction of quantum mechanics which is noncontextual, a point emphasized by various authors \cite{Spekkens_general_context},\cite{Mermin}.  Once an ontic state is specified, its response to experimental probes is often assumed to be independent of the quantum state, since the latter is just a statistical mixture of the former. This implicit assumption is called by Spekkens \cite{Spekkens_PBR}$\lambda$-sufficiency ($\lambda$ labels ontic states). Given this assumption, the conspiracy of measurement contextuality looks even more contrived as one needs to imagine that an arbitrary quantum state whose distribution in ontic space may be rather arbitrary (subject to constraints by quantum mechanics) always manages to yield the same probability for a projective experiment outcome despite the contextual response of the underlying ontic states which is, on the other hand, independent of the quantum state. I shall elaborate on this in the following sections. \\

In this letter, I would like to analyze how an ontological model conspires with measurement contextuality so as to comply with quantum mechanical predictions by attempts to describe how an ontic state is disturbed by measurements in different contexts. I argue strongly on the implausibility of the $\lambda$-sufficiency assumption for ontological models with measurement contextuality (some ontological models are not measurement contextual, more discussions in section \ref{lambda_sufficiency} and \ref{Discussions}). In section \ref{Ontological space}, I analyze the general feature of an ontic space that is related to experiment measurements, giving an intuitive picture of how an ontic state responds to the measurements. In section \ref{Measurement contextuality}, I compare a projective measurement under different contexts to analyze how could ontic states respond to projective measurements in different contexts and yield correct noncontextual quantum mechanical predictions. In section \ref{lambda_sufficiency}, I discuss the implication of $\lambda$-sufficiency to measurement contextuality. Finally, in section \ref{Discussions}, I summarize the results, and conclude with a discussion of the ontological picture put forward in the present letter in the context of several ontological models and no-go theorems in the literature. \\

\section{Ontological space} \label{Ontological space}

Before describing the ontological space, it is necessary to first list the notations. I'll call  the system subject to experimental probes S, the pure quantum state in which we prepare the system is denoted by $\psi$. The ontological space underlying the system is called $\Lambda$ and an individual ontic state in the space is called $\lambda$. A general POVM is denoted by $\{E_i\}$, a projective measurement will be written as $\prod_E$. A rank-1 projective measurement projecting onto a pure quantum state $\psi$ will be denoted by $\prod_{E_\psi}$; if the quantum state is labeled by an index (used to denote a member of an orthonormal basis), say $\psi_i$, the corresponding projective measurement will be denoted by $\prod_{E_i}$. The ontic distribution of $\psi$ over the ontological space is expressed as $\rho(\lambda|\psi)$ and the probability of an ontic state responding to a given experiment outcome $E_i$ will be encoded by an indicator function $\xi_{E_i}(\lambda)$. In general, the preparation and measurement setting could affect $\rho(\lambda|\psi)$ and $\xi_{E_i}(\lambda)$ respectively and I'll denote such dependence as $\rho(\lambda|\psi, P)$ and $\xi_{E_i}(\lambda, M)$ accordingly where $P$ and $M$ refers to preparation and measurement setting respectively. I shall write the support of $\rho$ and $\xi$ as $\mathrm{Supp}(\rho(\lambda|\psi,P))$ and $\mathrm{Supp}(\xi_{E_i}(\lambda,M))$ respectively.  In the absence of such labels, they correspond to the support of the union of all the possible settings.\\

Quantum mechanics puts constraints on possible ontic state properties and I'll demonstrate in this and the next section that in fact some constraints are rather strong. According to Born's rule, the probability of getting experiment outcome $E_i$ for a POVM on a quantum state $\psi$ is 

\begin{eqnarray}
P_{\psi}(E_i)=\langle\psi|E_i|\psi\rangle, \label{Born}
\end{eqnarray}
where $E_i\geq0$ and $\sum_iE_i=1$. The Born's rule is noncontextual in that given the quantum state and the measurement operator, the probability $P_{\psi}(E_i)$ is fixed, independent of settings. In the language of ontological models, the  probability can be written as

\begin{eqnarray}
P_{\psi}(E_i)=\int d\lambda \hspace{2pt}\xi_{E_i}(\lambda, M)\rho(\lambda|\psi,P), \label{ont_expt}
\end{eqnarray}
where I have simplified the notation for the measure over the ontic states as $d\lambda$ and proper integral measure is implied. The above expression clearly shows the constraint quantum mechanical prediction puts on $\lambda$ as different $M$ and $P$ need to conform to the same probability. Given that the indicator function is independent of the quantum state (i.e., $\lambda$-sufficiency in the Introduction), the constraint by equation (\ref{ont_expt}) is extremely strong. To illustrate this point, let's imagine the possible scenarios in which it may be satisfied. The simplest way the constraint can be satisfied is to prepare a quantum state whose ontic states are distributed uniformly over its support. For instance, in the discrete case, it is only nonzero for $\lambda_i, i=1,\cdots, n;$ so that $\rho(\lambda|\psi,P)=1/n$. Consider a compatible set of projective measurements in Hilbert space dimension d larger than 2, i.e, $\{\prod_{E_i}, i=1,\cdots, d\geq3\}$. Assume the response of an ontic state to the measurements is sharp (outcome deterministic), i.e., $\xi^2_{E_i}(\lambda,M)=\xi_{E_i}(\lambda,M)$. Due to measurement contextuality, the support of $\xi_{E_i}(\lambda,M)$ may differ depending on $M$. To satisfy equation (\ref{ont_expt}) for different $\mathrm{Supp}(\xi_{E_i}(\lambda,M))$, one needs to demand that the overlap area between the support of the indicator function and that of the quantum state remains constant since $\rho$ is uniform over its support. This requirement looks rather easy to meet. However, as the constraint applies to an arbitrary quantum state whose distribution over ontic space is supposed to differ from one quantum state to another in many (possibly infinite) different ways, it looks extremely (if not impossible) difficult to accommodate all these variations with the fixed indicator functions for each of the say, two different measurement settings $M$ and $M'$ corresponding to the same projective measurement $\prod_{E_i}$. In the next section, I shall give a more quantitative argument once we have a more quantitative picture of the structure of the ontological space and its response to measurements. \\

Let's now proceed to find out what general properties an ontological space needs to satisfy related to quantum states and measurements. The normalization condition $\sum_i\xi_{E_i}(\lambda,M)=1$ says that the total probability of the response of an ontic state to all possible measurement outcomes of a POVM is equal to one. From this fact, we derive the first property of the ontological space relative to measurements: \\

\begin{lemma}
\label{indicator space} For a quantum system S, the support of the set of indicator functions  is invariant with respect to POVM. \\
\end{lemma}

Proof. We need to prove that $\forall$ $\{E_i\}\neq\{E'_i\}$, $\cup_i\mathrm{Supp}(\xi_{E_i}(\lambda))=\cup_i\mathrm{Supp} (\xi_{E'_i}(\lambda))$, where $\{E\}$ and $\{E'\}$ refer to any two distinct POVM. This is easily proved by ad absurdum. Suppose $\exists \lambda_0$ such that $\lambda_0\in\cup_i\mathrm{Supp}(\xi_{E_i}(\lambda)$ and $\lambda_0\notin\cup_i\mathrm{Supp}(\xi_{E'_i}(\lambda))$, then $\sum_i\xi_{E'_i}(\lambda_0)=0$. Contradiction. $\square$ \\

From this Lemma, we see that the ontological space for a quantum system is fixed in the sense that it is always the same set of ontic states that can respond to different sets of measurements and I shall call the invariant ontic set $\Lambda_{\xi}$. Orthogonality condition for mutually orthogonal quantum states allows $\Lambda_{\xi}$ to be divided into subsets corresponding to members of a complete set of jointly performed projective measurements. This gives the following Lemma: \\

\begin{lemma}
\label{orthogonality} $\forall \prod_{E_i}\cdot\prod_{E_j}=0$ and $\sum_i\prod_{E_i}=1$, $\mathrm{Supp}(\xi_{E_i}(\lambda,M))\cap\mathrm{Supp}(\rho(\lambda|\psi_j))=0$        \textnormal{(}more precisely, zero measure\textnormal{)}, where $\prod_{E_i}|\psi_j\rangle=\delta_{ij}|\psi_j\rangle=0$.
\end{lemma}
Proof. By ad absurdum. If $\mathrm{Supp}(\xi_{E_i}(\lambda,M))\cap\mathrm{Supp}(\rho(\lambda|\psi_j)) (i\neq j)$ has nonzero measure, then for the quantum states $\psi_j$, the probability of getting outcome $E_i$ is nonzero. Contradiction. $\square$ \\

For outcome deterministic projective measurements, we have the following stronger result: \\
\begin{lemma}
\label{deterministic measurements} $\forall$ outcome deterministic projective measurements: $\prod_{E_i}\cdot\prod_{E_j}=0$ and $\sum_i\prod_{E_i}=1$, $\mathrm{Supp}(\xi_{E_i}(\lambda,M))\cap\mathrm{Supp}(\xi_{E_j}(\lambda,M))=0$. \\
\end{lemma}

Proof. The normalization condition $\sum_i\xi_{E_i}(\lambda,M)=1$ together with $\xi^2_{E_i}(\lambda,M)=\xi_{E_i}(\lambda,M)$ imply the disjointness for mutually orthogonal projective measurements. $\square$ \\

The measurement contextuality implies that $\exists$ $|\psi\rangle$ and $M$, $\mathrm{Supp}(\rho(\lambda|\psi))\subset\mathrm{Supp}(\xi_{E_\psi}(\lambda,M))$, where $M$ is a certain measurement setting, $E_\psi=|\psi\rangle\langle\psi|$. This property is first derived in \cite{Rudolph_Harrigan} in which they name it deficiency (note that it also applies to outcome indeterministic measurements where measurement contextuality doesn't necessarily hold). Here I generalize it a little bit by including all possible preparations for $|\psi\rangle$ in defining the support of $\psi$ and I shall call the support space $\Lambda_\psi$. This property says that for at least one projective measurement setting, the support of the indicator function (also denoted for notation simplicity as $\Lambda_{\xi_i,M}$) is strictly larger than that of the corresponding quantum state. \\

Now suppose the projective measurement to which some of the ontic states are contextual is $\prod_{E_{i_0}}$ and the rest of the jointly measured projective measurements are given by $\prod_{E_i}$ ($i\neq i_0)$. The corresponding quantum states $|\psi_{i_0}\rangle$ and $|\psi_i\rangle$ ($i\neq i_0$) form an orthonormal basis. Depending on the measurement settings, the set of orthonormal states $|\psi_i\rangle$ ($i\neq i_0$) are rotated unitarily from one set to another in the subspace $1-\prod_{E_{i_0}}$. Due to Lemma 2, any ontic state that belongs to the support of any quantum state in the subspace $1-\prod_{E_{i_0}}$ is noncontextual to $\prod_{E_{i_0}}$. On the other hand, any ontic state that belongs to the support of $|\psi_{i_0}\rangle$ has to be noncontextual to $\prod_{E_{i_0}}$ too, due to normalization condition of the quantum state $\psi_{i_0}$. So we see that only ontic states that are outside the support of $1-\prod_{E_{i_0}}$ and $|\psi_{i_0}\rangle$ can be contextual to $\prod_{E_{i_0}}$ (This conclusion is valid regardless of outcome determinacy). \\

Recall that measurement contextuality (for outcome deterministic measurements) requires that any ontic state must be contextual to some projective measurement. This combined with the conclusion of the above paragraph gives the following corollary: \\

\begin{corollary}
\label{contextual measurements} There exists more than one measurement outcome that are contextual for outcome deterministic measurements. \\
\end{corollary}

\begin{figure}[h!]
\begin{center}
\includegraphics[width=0.4\textwidth]{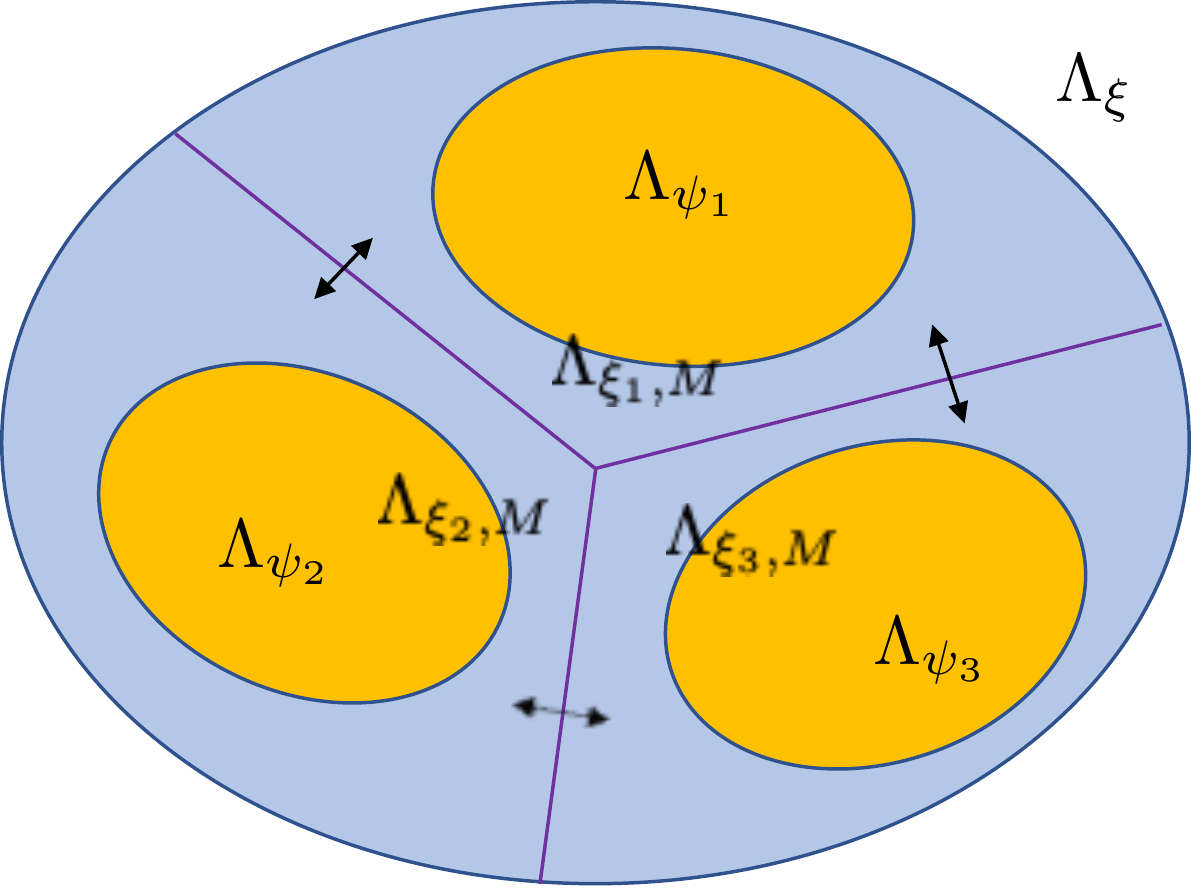}
\end{center}
\caption{A sketch of the ontological space. The ontological space $\Lambda_\xi$ encloses the whole oval region which is invariant to different POVM. The three disjoint oval regions (in yellow) correspond to $\Lambda_{\psi_i}$ - ontological spaces of orthonormal quantum states $\psi_i$, $i=1,2,3$. For outcome deterministic projective measurements, the the entire space can be divided into disjoint subregions (delineated by purple lines) each of which corresponds to the support of the indicator function of the projective measurement $\prod_{E_i}=|\psi_i\rangle\langle\psi_i|, i=1,2,3$. Due to deficiency, at least one of the regions $\Lambda_{\xi_i,M}$ is strictly larger than the corresponding $\Lambda_{\psi_i}$. Changing measurement settings may shift the boundaries of the supports of the indicator function for the corresponding measurement (any of the $\prod_{E_i}$) subject to different contexts, as indicated by double-headed arrows on the purple lines, but they never cross the boundaries of $\cup_i\Lambda_{\psi_i}$. For example, when changing measurement settings for $\prod_{E_1}$, the boundary of $\Lambda_{\xi_1,M}$ can move but never crosses any of $\Lambda_{\psi_i}$. The same applies to other projective measurements onto $\psi_i$, $i\neq1$. For outcome indeterministic measurements, the overall picture remains valid with the exception that distinct $\Lambda_{\xi_i,M}$ can overlap. Figure 2 of \cite{Rudolph_qutrit} provides a concrete example of ontic state dependence on measurement contexts.}\label{fig_ontic_space}
\vspace{10pt}
\end{figure}

Lemma \ref{indicator space} \textendash \hspace{2pt} \ref{deterministic measurements} together with deficiency 
discussed above give us an overall picture of the ontological space underlying a quantum system. Geometrically, the ontological space for a quantum system can be partitioned according to an orthonormal set of quantum states $\psi_i$ and the corresponding projective measurements $\prod_{E_i}$ sketched in Figure \ref{fig_ontic_space}. Different POVM (including projective measurements of course) leave the boundary of the ontological space $\Lambda_\xi$ (defined below the proof of Lemma \ref{indicator space}) invariant. Different projective measurement settings for any of the projective measurement that belongs to $\prod_{E_i}, i=1,\cdots,d$ ($d$ is Hilbert sapce dimension) can only affect ontic states that lie outside the support of the corresponding orthonormal basis, i.e., $\lambda_c\in \Lambda_\xi-\cup_i\Lambda_{\psi_i}$ ($\lambda_c$ denotes ontic states that are contextual to measurement settings and $\Lambda_{\psi_i}$ is defined in the paragraph below the proof of Lemma \ref{deterministic measurements}), such that the boundaries between the supports of the indicator functions of distinct members of jointly performed projective measurements $\prod_{E_i}$ can move due to change of measurement settings but they never cross the boundaries of $\cup_i\Lambda_{\psi_i}$ (for outcome indeterministic measurements, supports of indicator functions can have overlaps). \\

\section{Measurement contextuality} \label{Measurement contextuality}

We now discuss in more details how to satisfy the constraint of noncontextual quantum mechanical probabilities for projective measurement outcomes. Consider a quantum system S with orthonormal basis $\{\psi_i\}$ and prepare it in (in setting P) a quantum state $\psi$ given by
\begin{eqnarray}
|\psi\rangle=\sum_ia_i|\psi_i\rangle, \label{gs}
\end{eqnarray}
where $\sum_i|a_i|^2=1$. We first perform projective measurement $M$ with $\prod_{E_i}=|\psi_i\rangle\langle\psi_i|$ and $\sum_i\prod_{E_i}=1$. The probability of getting outcome say $E_1$ is \\

\begin{eqnarray}
P_\psi(E_1)=|a_1|^2=\int d\lambda \hspace{2pt} \xi_{E_1}(\lambda,M)\rho(\lambda|\psi,P). \label{measure_context1}
\end{eqnarray}

For now we assume outcome determinacy, i.e., $\xi^2_{E_i}(\lambda,M)=\xi_{E_i}(\lambda,M)$ and the above integral becomes integral over the support of the indicator function $\xi_{E_1}(\lambda,M)$.  Without loss of generality, we assume that the projective measurement $\prod_{E_1}$ is contextual and perform next a different projective measurement $M'$ on the same quantum state $\psi$ where $E_1=|\psi_1\rangle\langle\psi_1|$ is the same as given by $M$ but the rest of the measurements differ. The probability of getting $E_1$ is the same and can be similarly written as \\

\begin{eqnarray}
P_\psi(E_1)=|a_1|^2=\int d\lambda \hspace{2pt} \xi_{E_1}(\lambda,M')\rho(\lambda|\psi,P). \label{measure_context2}
\end{eqnarray}

Comparing equation (\ref{measure_context1}) and (\ref{measure_context2}), together with the outcome determinacy, we arrive at the following constraint \\

\begin{eqnarray}
 \int \underset{{\mathrm{Supp}(\xi_{E_1}(\lambda,M))}}{d\lambda} \hspace{2pt}\rho(\lambda|\psi,P)=  \int \underset{ {\mathrm{Supp}(\xi_{E_1}(\lambda,M'))}}{d\lambda}  \hspace{2pt} \rho(\lambda|\psi,P). \label{measure_context}
\end{eqnarray}
This is nothing but the condition discussed below equation (\ref{ont_expt}). As argued there, it is extremely contrived to satisfy equation (\ref{measure_context})  with two different supports independent of the distribution over the ontic states of any quantum state $\psi$. We could prepare infinite many different quantum states and subject each of them to the same two sets of projective measurements $M$ and $M'$ and equation (\ref{measure_context}) has to be satisfied for each quantum state. \\

Recalling the physical picture of the ontological space we obtained in the previous section, in order to satisfy equation (\ref{measure_context}), the only obvious possibility (that is not contrived) is such that $\mathrm{Supp}(\rho(\lambda|\psi,P))\subseteq\cup_i\Lambda_{\psi_i}$ (recall that $\cup_i\Lambda_{\psi_i}$ is noncontextual to $\prod_{E_1}$). However, $\cup_i\Lambda_{\psi_i}\subset\Lambda_\xi$ and is fixed, which implies that the requirement $\mathrm{Supp}(\rho(\lambda|\psi,P))\subseteq\cup_i\Lambda_{\psi_i}$ is impossible to satisfy. This is because the requirement would imply that there are regions of ontological space $\Lambda_\xi-\cup_i\Lambda_{\psi_i}$ that are support of no quantum state whatsoever! \\

So we are left with the contrived scenario in which $\rho(\lambda|\psi,P)$ needs to distribute over ontic states in a contrived way to always satisfy equation (\ref{measure_context}). It's even more contrived if one considers performing all other possible projective measurement settings besides $M$ and $M'$ for which same condition as equation (\ref{measure_context}) needs to be satisfied with the same $\rho(\lambda|\psi,P)$. The lesson we learned from this analysis is that the ontic space distribution of a quantum state can NOT be 'ergodic' over the full ontological space as we change continuously the quantum state throughout the whole Hilbert space. On the contrary, the distribution over ontic space has to be highly restricted and contrived so as to comply with noncontextual measurement outcome statistics. This is of course not inconceivable, for instance, the toy model by Spekkens \cite{Spekkens_epi} puts constraint on the maximum information of any quantum state can have, thus constraining its support on ontological space.  Clearly one needs more constraints in order to reconcile measurement contextuality with quantum mechanical noncontextuality (see also discussion concerning the ontic space distributions of quantum states in \cite{Rudolph_qutrit}). \\

So what are the possible scenarios to satisfy measurement contextuality? An option is that the supports of the indicator functions in different settings only differ by zero measure. This looks rather pathological for a well-behaved ontological model. So we restrict to cases with finite measure difference over ontic states. It is hard to proceed without any further constraint on the underlying ontological models. A reasonable assumption is that the ontological model should at least inherit some of the properties of quantum mechanical structure so as to be consistent with quantum mechanical predictions. One pertinent property to measurements is that different projective measurement settings can be obtained from each other by unitary rotations in Hilbert space. If the underlying ontic states respond to them in a similar way (for instance, the ontological qutrit models in \cite{Rudolph_qutrit}), then it is reasonable to expect that once we specify the indicator functions in all different settings for one projective measurement, say $\prod_{E_1}$ discussed here, we should be able to obtain all other indicator functions for all projective measurements by unitary-like symmetry operations whose quantum mechanical counterpart acts to rotate in Hilbert space $\psi_1$ to any pure quantum state which the new projective measurement project onto. Recall the picture of the ontological space (cf. Figure \ref{fig_ontic_space}) obtained in the previous section, only the interstitial region $\Lambda_\xi-\cup_i\Lambda_{\psi_i}$ can be contextual to each of the projective measurements $\prod_{E_i}$. Over such a region, the supports of indicator functions can differ from each other and a given quantum state needs to satisfy equation (\ref{measure_context}) for any pair of them. The only plausible way to satisfy this is to require the ontic distribution of the given quantum state to be uniform over the region $\cup_{\{M\}}\mathrm{Supp}(\xi_{E_i}(\lambda,M))-\cap_{\{M\}}\mathrm{Supp}(\xi_{E_i}(\lambda,M))$ for a measurement $\prod_{E_i}$ (to which we consider contextuality) where $\{M\}$ refers to the complete set of contexts keeping $\prod_{E_i}$ fixed (with further requirements such as incompressibility of $\Lambda_{\xi_i,M}$). Now in light of the symmetry considerations, the interstitial region is 'rotated' continuously according to the continuous rotation of the projective measurements to which we probe contextuality, i.e., when we consider contexutality to a new projective measurement, say $\prod_{E_{1'}}$, where $|\psi_{1'}\rangle\neq |\psi_1\rangle$, the whole experiment settings for the new projective measurement can be obtained from those with respect to $\prod_{E_1}$ by unitary rotations in the Hilbert space. Since any ontic state needs to be contextual to some projective measurement, the interstitial region needs to be ergodic over the whole space $\Lambda_\xi$! The ergodicity and the continuity of the interstitial region movement in the ontological space in response to unitary transformations of projective measurements in the Hilbert space essentially imply that for the quantum state to comply with all possible measurement predictions, it needs to have uniform support over the whole ontological space $\Lambda_\xi$, a clear contradiction. By the similar token, it is reasonable to require the ontological distribution of the quantum state to be rigidly transported with the unitary transformation of the quantum state in the Hilbert space (analogous to phase space representations of quantum states, see e.g. \cite{Ferrie}). As the distribution is transported continuously in the ontological space, it becomes impossible to always satisfy equation (\ref{measure_context}) for any two distinct supports of $\xi_{E_1}(\lambda, M)$ and $\xi_{E_1}(\lambda,M')$ (barring pathological case of zero measure difference). \\

A nice example to which the above arguments apply is the ontolgogical qutrit models given in \cite{Rudolph_qutrit}. In those models, both the ontic state distribution of quantum states and the ontic state response to measurements clearly satisfy the discussed symmetry with respective to quantum state/measurement transformation and so those models can not reproduce the correct measurement contextuality. \\

\section{$\lambda$-sufficiency vs. measurement contextuality} \label{lambda_sufficiency}

From the previous discussion, it is clear that the difficulty in satisfying the quantum mechanical noncontextual predictions with measurement contextual ontic states lies in the assumption of $\lambda$-sufficiency. This requirement fixes the response of ontic states given measurement settings and the response is independent of quantum states subject to the measurements. This is perhaps not so surprising in the sense that $\lambda$-sufficiency is a classical notion (a quantum state being viewed as a statistical mixture over ontic states) and measurement contextuality is non-classical still beyond our classical reasoning. At this point, one can't help speculating that $\lambda$-sufficiency may not be the right assumption  to start with an ontological model. Perhaps in a more plausible ontological model, it is desirable to build in some ways from the beginning a physically motivated connection of a quantum state with its underlying ontic states' response to measurements. I therefore propose the following conjecture: \\

\begin{conjecture}
\label{lambda sufficiency} For ontological models with measurement contextuality, $\lambda$-sufficiency can not be true if the underlying model has certain symmetry with respect to projective measurements such as the rotational symmetry of the ontic distribution of quantum state or/and of the response of the ontic states in the ontological space according to quantum state transformation or/and projective measurement transformation in Hilbert space respectively. \\
\end{conjecture}

The ontological qutrit models in \cite{Rudolph_qutrit} is an example consistent with the above conjecture. In the discussion of section \ref{Measurement contextuality}, I have assumed outcome determinacy for projective measurements. A natural question is what may change if the assumption is relaxed  to allow indetermincay. For outcome indeterministic projective measurements, measurement contextuality may not be deduced by itself (such as in certain two dimensional ontological models discussed in \cite{Spekkens_general_context} for generalized measurement contextuality). For example, for trivial ontological models taking a quantum mechanical wave function itself as the ontic state (it belongs to $\psi$-complete models in the terminology of \cite{Harrigan_Spekkens}), we have preparation contextuality (in the sense of \cite{Spekkens_general_context}) but not measurement contextuality (for the trivial reason that they essentially reduce to quantum mechanical formalism). On the other hand, Lemma \ref{indicator space}, \ref{orthogonality} and deficiency discussed in section \ref{Ontological space} still hold. The intuitive picture of the ontological space obtained previously remains the same with the exception that now indicator functions corresponding to distinct projective measurements that are jointly performed can have overlaps in their supports of ontic states. For an ontological model with measurement contextuality, the argument in the previous section still applies and we have to conclude that quantum states have to conspire with measurement contextuality in highly nontrivial ways to comply with quantum mechanical predictions\cite{comment_indeterminacy}. \\

\section{Discussions}\label{Discussions}

Through the above sections, we have obtained an intuitive picture of the ontological space in response to measurements. It is invariant to different POVM (Lemma \ref{indicator space}) and the ontic space of the indicator functions of jointly performed projective measurements  has no intersection with that of the corresponding eigenstates orthogonal to the projected subspace by the measurements (Lemma \ref{orthogonality}). In the case of outcome deterministic measurements, the ontic space of the indicator functions themselves are mutually disjoint (Lemma \ref{deterministic measurements}). Due to deficiency, the ontic space of at least one indicator function is strictly larger than the corresponding eigenstate. In measurement contextual ontological models, only ontic states belonging to $\Lambda_\xi-\cup_i\Lambda_{\psi_i}$ can be affected by changing settings of projective measurements while keeping one of $\prod_{E_i}$ unchanged. This is true regardless of the outcome determinacy of a projective measurement. \\ 

We have also argued that $\lambda$-sufficiency makes measurement contextuality unreasonably difficult to comply with quantum mechanical noncontextual predictions. We further conjecture that any measurement contextual ontological model with certain symmetry such as rotation symmetry in the ontological space with respect to quantum states or/and projective measurements has to abandon $\lambda$-sufficiency. This may be part of the reason why a successful $\psi$-epistemic \cite{Harrigan_Spekkens} ontological model in dimension larger than two is so difficult to construct (see the ontological models of qutrit in \cite{Rudolph_qutrit} that approximate quantum mechanical predictions). On the other hand, the highly artificial ontological model of Bell\cite{Bell} satisfies measurement contextuality by requiring the response of ontic states to measurements to explicitly depend on the quantum state, violating $\lambda$-sufficiency. The $\psi$-epistemic models constructed in \cite{Rudolph} share the same feature. In this aspect, the hidden-variable model of Bohm \cite{Bohm} (it belongs to the category of $\psi$-supplemented models in \cite{Harrigan_Spekkens}, which together with the other two \textendash \hspace{2pt} $\psi$-complete and $\psi$-epistemic already discussed constitute the whole space of ontological models) satisfies measurement contextuality by requiring the behavior of particle position (part of the ontic state in the language of ontological approach) to depend on its quantum state explicitly. From the perspective of $\lambda$-sufficiency, the response of ontic states in the aforementioned models seems to depend on quantum states in a too ad hoc \cite{Bell},\cite{Rudolph} or strong \cite{Bohm} way so as to reproduce quantum mechanical predictions of measurements in general Hilbert space dimensions. In the present author's opinion, ontological models that manage to balance the dependence on ontic states and on quantum states in response to measurements look more plausible to be considered as part of the successful intuitively and physically-motivated framework underlying quantum mechanics. For example, some bigger symmetry combining quantum state ontic distribution with responses of indicator functions appears to be a good property worth seeking. At this point, it's worth mentioning that $\lambda$-sufficiency plays a crucial role in various no-go theorems concerning the reality of quantum states. Both the PBR \cite{PBR} and Hardy \cite{Hardy} argument on the reality of quantum states assume $\lambda$-sufficiency. If $\lambda$-sufficiency were to go, then we need not require quantum state to be realistic (cf. \cite{Rudolph}). Again modifying $\lambda$-sufficiency seems a good option for constructing successful $\psi$-epistemic ontological models to address measurement contextuality. \\

In this letter, we have studied measurement contextuality in ontological models in general terms. Detailed criteria for Conjecture \ref{lambda sufficiency} would require constructing concrete ontological models and classifying them systematically. The quantitative symmetry properties of ontological models with respect to quantum states and measurements and rigorous proof of Conjecture \ref{lambda sufficiency} are left for future work. Finally, possibilities of measurement affecting ontic state distribution  (retrocausality) is another option worth further considerations. \\ 

\section*{Acknowledgement}

This work was supported by the start-up funds from Texas A\&M University.


\begin{thebibliography}{99}
\bibitem{Spekkens_epi}Robert W. Spekkens, Phys. Rev. A, 75, 032110 (2007)
\bibitem{Spekkens and etc} Robert W. Spekkens Pirsa: 08020051 - Why the quantum? Insights from classical theories with a statistical restriction
\bibitem{Rudolph_Harrigan} Nicholas Harrigan and Terry Rudolph, arXiv:0709.4266
\bibitem{Bell} John S. Bell, Rev. Mod. Phys., 38, 3, 447, 1966
\bibitem{KS} S. Kochen and E. P. Specker, J. Math. Mech. 17, 59, 1967
\bibitem{Spekkens_general_context} Robert W. Spekkens, Phys. Rev. A, 71, 052108 (2005)                 
\bibitem{Mermin} N. David Mermin, Rev. Mod. Phys., 65, 3, 803, 1993
\bibitem{Spekkens_PBR} Robert W. Spekkens, Pirsa: 12050021 - Why I Am Not a Psi-ontologist
\bibitem{Rudolph_qutrit} Terry Rudolph, arXiv:quant-ph/0608120

\bibitem{Ferrie} Christopher Ferrie and Joseph Emerson, J. Phys. A: Math. Theor. 41 352001 (2008)
\bibitem{Harrigan_Spekkens} Nicholas Harrigan and Robert W. Spekkens, Found Phys (2010) 40: 125–157
\bibitem{comment_indeterminacy} In the case of outcome indeterminacy, there's not a proof that any ontic state has to be contextual to some measurements if the underlying model is measurement contextual. If not every ontic state has to be contextual, then the ergodicity of the interstitial region for outcome deterministic case in section \ref{Measurement contextuality} no longer holds. But I bet Conjecture \ref{lambda sufficiency} is still true regardless of outcome determinacy.
\bibitem{Rudolph} Peter G. Lewis, David Jennings, Jonathan Barrett, and Terry Rudolph, Phys. Rev. Lett. 109, 150404, 2012
\bibitem{Bohm} David Bohm, Phys. Rev., 85, 166, 1952.
\bibitem{PBR} Matthew F. Pusey, Jonathan Barrett and Terry Rudolph, Nature Physics 8, 475-478 (2012)
\bibitem{Hardy} Lucien Hardy, arXiv:1205.1439

\end{thebibliography}
\end{document}